\documentclass[
  prb,
  twocolumn,
  preprintnumbers,
  floatfix,
  showpacs,
  amsmath,
  amssymb,
]{revtex4-1}

\usepackage[final]{graphicx}
\usepackage{tabularx}
\usepackage[nooneline]{subfigure}
\usepackage{amsmath,amssymb}
\usepackage{dcolumn}
\usepackage{color}
\usepackage{times}

\newcommand{\etal}[0]{\textit{et al.}}

\newcommand{\K}[0]{\text{K}}

\newcommand{\eV}[0]{\text{eV}}

\newcommand{\atom}[0]{\text{atom}}

\newcommand{\Fe}[0]{\text{Fe}}
\newcommand{\Cr}[0]{\text{Cr}}
\newcommand{\tot}[0]{\text{tot}}
\newcommand{\sol}[0]{\text{sol}}

\newcommand{\sect}[1]{Sect.~\ref{#1}}
\newcommand{\fig}[1]{Fig.~\ref{#1}}
\newcommand{\eq}[1]{Eq.~(\ref{#1})}

\renewcommand{\epsilon}[0]{\varepsilon}

\begin{document}

\preprint{
  published as Phys. Rev. B {\bf 77}, 134206 (2008) \hspace{156pt}
  Copyright (2008) The American Physical Society
}

\title{
  Short-range order and precipitation in Fe-rich Fe--Cr alloys: \\
  Atomistic off-lattice Monte Carlo simulations
}

\date{December 17, 2007}

\author{Paul Erhart}
\email{erhart1@llnl.gov}
\author{Alfredo Caro}
\author{Magdalena Serrano de Caro}
\author{Babak Sadigh}
\affiliation{
  Chemistry, Materials, Environmental, and Life Sciences Directorate,
  Lawrence Livermore National Laboratory,
  L-367, Livermore, California, 94550
}


\begin{abstract}
Short-range order (SRO) in Fe-rich Fe--Cr alloys is investigated by
means of atomistic off-lattice Monte Carlo simulations in the
semi-grand canonical ensemble using classical interatomic
potentials. The SRO parameter defined by Cowley [Phys. Rev. B {\bf 77}, 669 (1950)] is used to
quantify the degree of ordering. In agreement with experiments a strong
ordering tendency in the Cr distribution at low Cr concentrations
($\lesssim\,5\%$) is observed, as manifested in negative values of the
SRO parameters. For intermediate Cr concentrations
($5\%\lesssim\,c_{\Cr}\lesssim\,15\%$) the SRO parameter for
the $\alpha$-phase goes through a minimum, but at the solubility limit the
$\alpha$-phase still displays a rather strong SRO. In thermodynamic
equilibrium for concentrations within the two-phase region the SRO
parameter measured {\em over the entire} sample therefore comprises
the contributions from both the $\alpha$ and $\alpha'$-phases. If both
of these contributions are taken into account, it is possible to
quantitatively reproduce the experimental results and interpret their
physical implications. It is thereby shown that the inversion of the
SRO observed experimentally is due to the formation of stable
(supercritical) $\alpha'$-precipitates. It is not related to the loss
of SRO in the $\alpha$-phase or to the presence of unstable
(subcritical) Cr precipitates in the $\alpha$-phase.
\end{abstract}


\maketitle

\section{Introduction}

Iron-chromium steels are of great technological importance due to their
superior properties at high temperatures and under aggressive chemical
conditions. In particular, they are used in reactor environments because of
their good swelling and corrosion resistance.\cite{GarTolSen00, HisKohKlue98}
The precise origin of many of these beneficial features is still
uncertain. Creep resistance is customarily associated with the presence of
{\em small} $\alpha'$-precipitates, while embrittlement has been related to
the presence of {\em larger} $\alpha'$-precipitates. The $\alpha$ and
$\alpha'$-phases are the Fe-rich and Cr-rich solid solutions into which the
body-centered cubic (bcc) phase of the alloy decomposes at temperatures below
about 1000\,K. At yet higher temperatures a third phase appears around
equiatomic composition, known as the $\sigma$-phase.

The phase diagram for this alloy has been assessed using the CALPHAD
methodology.\cite{calphad, Din91} In this approach the alloy was assumed to
behave as a subregular solution at temperatures below the existence
range of the $\sigma$-phase. In other words, the heat of formation is
assumed to be positive over the entire composition and it is
parameterized by a second degree polynomial in the Redlich-Kister
expansion. There is, however, an ensemble of experimental observations
at relatively low temperatures in both annealed and irradiated
samples, which suggest a more complex behavior. \cite{KuwHam88,
  Sagkosarb01, ShaNikMuk01}

A breakthrough in the understanding of the microstructure of these alloys was
made through neutron diffraction measurements \cite{MirHenPar84} which showed
(1)
negative Cowley short-range order (SRO) parameters 
(see \eq{eq:sro} below) at small Cr concentrations indicating
strong short-range ordering of Cr atoms, and
(2)
an inversion of the sign of these parameters with increasing Cr
concentration, which seems to imply a change of SRO in the solid
solution as it approaches the solubility limit.
The implications of these discoveries for the interpretation of
creep resistance and embrittlement are apparent, since SRO is
known to affect the mobility of dislocations \cite{RodRodPro04} and
precipitation plays a role in intergranular cracking.\cite{KluHar01}

Nonetheless the understanding of the atomistic details of the complex behavior
of this alloy is still incomplete. This situation has motivated a number of
recent first-principles studies \cite{Hen83, OlsAbrWal06b, MirYalMir04,
  KlaDraFin06, ErhSadCar08} which addressed the energetics of this
system. These calculations revealed a change of sign in the heat of
formation of the solution from the negative to the positive side as
the Cr concentration increases above approximately $10\%$.\cite{Hen83,
  OlsAbrWal06b, MirYalMir04, KlaDraFin06, ErhSadCar08} Detailed
analysis of the origin of this anomaly showed that, while the heat of
solution of a Cr impurity in Fe is large and negative, magnetic
frustration leads to a strong Cr--Cr repulsion, causing the heat of
formation to assume large positive values as the Cr content
increases.

These advances in the understanding of the energetics of Fe--Cr alloys have
provided the basis for the development of accurate interatomic potential
models, which enable large scale simulations of the microstructural
evolution. Two approaches were developed known as the two band model
\cite{OlsWalDom05} and the composition dependent model, \cite{CarCroCar05}
which both address the complex shape of the heat of formation curve at 0\,K as
determined from the aforementioned first-principles calculations. A third
approach \cite{LavDraNgu07} to model the finite temperature properties
of this alloy is based on a cluster expansion fitted to density functional
theory data which is then used it in combination with lattice Monte
Carlo simulations to study precipitation and ordering. Using the
composition dependent model, we recently studied the implications of
the change in sign of the heat of formation on the thermodynamic
behavior at finite temperatures and proposed a modified phase diagram
for Fe--Cr in the region of low Cr content and temperatures below the
range of existence of the $\sigma$-phase.\cite{CarCarLop06b} In fact
compared to the CALPHAD assessment,\cite{Din91} which is based on the
assumption of a {\it standard segregating binary} mixture (positive
heat of mixing at all concentrations, compare above), a better
agreement with the experimental location of the \emph{solvus} was
obtained.

From our viewpoint three aspects render the study of SRO in Fe--Cr alloys
particularly interesting:
(1) Experimental measurements of SRO parameters of heterogeneous systems (such
as mixtures of the $\alpha$ and $\alpha'$-phases) provide only compound
quantities equivalent to averages over the entire sample.
(2) In Fe--Cr this situation is further complicated by the ordering tendency
of the system at low Cr concentrations.
(3) The recent advances in the development of Fe--Cr interatomic potentials
and the availability of suitable simulation techniques provide the possibility
to explore the atomistic details of these processes. Thereby, we are able to
resolve the contributions to the experimentally measured quantities and
understand the microscopic behavior of the material. We discuss the evolution
of the SRO parameter in a mixture of two phases and the evolution of the SRO
parameter upon coarsening, induced e.g., by aging or irradiation. It is shown
that the experimentally observed inversion of the SRO parameter is due to the
formation of stable (supercritical) $\alpha'$-precipitates. It is not related
to an overall loss of SRO, a loss of SRO in the $\alpha$-phase, or to the
presence of unstable (subcritical) Cr precipitates in the
$\alpha$-phase. Furthermore, by using the same order parameter as in the
experimental reports, we reconcile an apparent discrepancy between
experimental data and the model predictions reported in
Ref.~\onlinecite{CarCroCar05}.

The paper is organized as follows. Section \ref{sect:sro} introduces the order
parameters used in this work, while \sect{sect:method} describes the
methodology. Section \ref{sect:results} contains the results of the
simulations organized into three subsections, which address ({\it i}) the
relation between the chemical potential difference and the
concentration, ({\it ii}) the SRO in the $\alpha$-phase, and ({\it
  iii}) the evolution of the SRO in the two-phase region. In
\sect{sect:discussion} a simple expression for the SRO in the
two-phase region is introduced, which is subsequently employed to
interpret the experimental result and to derive the temperature
dependence of the SRO parameter. Section~\ref{sect:conclusions}
summarizes the conclusions.

\section{Measures of short-range order}
\label{sect:sro}

Short-range order in binary alloys is conveniently measured by means of the
order parameter introduced by Cowley \cite{Cow50} which has been widely
applied in the past (see e.g., Refs.~\onlinecite{OvcZviLit76, MirHenPar84,
  OvcGolGus06}). The SRO parameter for the $k$-th shell of a Cr atom is
defined as
\begin{align}
  \alpha_{\Cr}^{(k)} &= 1 - \frac{Z_{\Fe}^{(k)}}{Z_{\tot}^{(k)} (1-c_{\Cr})},
  \label{eq:sro}
\end{align}
where $Z_{\Fe}^{(k)}$ denotes the number of Fe atoms in the $k$-th shell,
$Z_{\tot}^{(k)}$ is the total number of atoms in the same shell and $c_{\Cr}$
is the overall (total) Cr concentration.
It is instructive to consider several special cases:
(1) If a Cr atom is surrounded exclusively by other Cr atoms,
$Z_{\Fe}^{(k)}=0$ and $\alpha_{\Cr}^{(k)}=1$.
(2) If one has a completely random distribution the ratio
$Z_{\Fe}^{(k)}/Z_{\tot}^{(k)}$ equals $c_{\Fe}=1-c_{\Cr}$ and
$\alpha_{\Cr}^{(k)}=0$.
(3) Finally, if each Cr atom is surrounded exclusively by Fe atoms
$Z_{\Fe}^{(k)}=Z_{\tot}^{(k)}$ and therefore
\begin{align}
  \alpha_{\Cr}^{(k)} &= -\frac{c_{\Cr}}{1-c_{\Cr}},
  \label{eq:sro_limit}
\end{align}
which provides a lower bound to the SRO parameter, in other words it indicates
the maximum possible degree of short-range order at a given composition.

Previous experimental studies of SRO in Fe--Cr have used a SRO parameter
specific for the bcc lattice.\cite{MirHenPar84} This parameter is defined as a
coordination number-weighted average of the SRO parameters defined above for
the first and second neighbor shells
\begin{align}
  \beta = \frac{8 \alpha_{\Cr}^{(1)} + 6 \alpha_{\Cr}^{(2)}}{14}.
  \label{eq:beta}
\end{align}
Other studies \cite{OvcGolGus06} assumed the SRO parameters
$\alpha_{\Cr}^{(1)}$ and $\alpha_{\Cr}^{(2)}$ to be the same which leads to
similar results. For consistency with literature and to simplify comparison we
focus on this bcc-specific weighted SRO parameter in the following
sections. In \sect{sect:sro_ss} we will however explicitly discuss the SRO
parameters for different shells, which yields insights into the range of
Cr--Cr interactions.

\section{Methodology}
\label{sect:method}

\subsection{Algorithm}
\label{sect:algorithm}

If one is to study the problem of short-range order and the
formation of precipitates in a segregating alloy while including the
effects of local relaxations, one requires an algorithm, which is
capable of efficiently taking into account both chemical mixing and
atomic relaxation. The former can be achieved by employing a
Monte-Carlo (MC) algorithm which allows changes in the chemical identity of
the particles as will be discussed in the following two paragraphs. In
principle atomic relaxations can be captured by a displacement MC
algorithm but for the problem at hand molecular dynamics (MD)
simulations turn out to be more efficient. The basic approach is thus
to alternate MC moves and MD steps.

The usual approach to model the effects of chemical mixing in
heterogeneous solid solutions, is to employ a MC swap algorithm, in
which the global composition is kept constant and the system is
advanced by swapping pairs of particles of different chemical
identities. The change in total energy is employed to decide whether a
move is accepted or rejected. For single phase systems one can also
use a semi-grand canonical ensemble (SGC), in which the composition is
determined by the difference of the chemical potentials $\Delta\mu$ of
the species, and each MC move is equivalent to exchanging particles
with the external reservoirs of both chemical species. Unlike the swap
scheme the SGC-MC algorithm can be formulated to operate only on one
particle at a time. Therefore, the latter can be more easily
parallelized.

Within a two-phase region of the phase diagram the relation between
the concentration $c$ and the chemical potential difference
$\Delta\mu$ has an infinite slope. In principle, the SGC-MC algorithm
cannot deal with such situations, and in order to keep the global
composition within the miscibility gap, one must apply an additional
constraint. The latter can be formally derived as a constraint on the
concentration fluctuations, the formal derivation of which will be
given elsewhere.\cite{mccask} For the present study, we implemented
the constraint in a very simple way, namely as a limitation on the
number of particles in the reservoirs: When the (externally specified)
maximum global concentration of solute atoms is reached, MC moves
which lead to a further increase of the solute concentration are
always rejected, regardless of their energy, i. e. the maximum number
of solute atoms is limited. To clarify this approach, consider the
case in which constraints are imposed on both the maximum and the
minimum number of solute atoms. In the (hypothetical) limit that the
difference between the lower and upper limit vanishes, this algorithm
would become equivalent to the swapping algorithm. We recall here that
the primary motivation behind the development of this algorithm is
that it can efficiently be parallelized.

To summarize, for values of $\Delta\mu$ within the $\alpha$-phase, our
simulations start with pure Fe samples and by alternating MC
moves with MD steps the sample loads solute atoms and relaxes the
lattice in the solid solution until the equilibrium composition is
reached and stationary fluctuations reach canonical values. If
$\Delta\mu$ exceeds the stability range of the $\alpha$-phase, the
simulation still starts from a pure Fe sample and first establishes a
solid solution in the $\alpha$-phase. Since the formation of a
supercritical nucleus is linked to a nucleation barrier, precipitation
does not occur immediately but only after some residence time. The
system therefore evolves initially in a single phase. However, as soon
as the spatial fluctuations in the composition have lead to the
formation of supercritical nuclei, they will grow rapidly and the
global Cr concentration rises sharply. When the target global
composition is reached, or equivalently, when the volume fraction of
$\alpha'$-precipitates reaches the target value, the concentration
constraint keeps the system fluctuating around the pre-set global
composition, and coarsening of the precipitates is observed. The
residence time prior to the formation of stable $\alpha'$-precipitates
depends on $\Delta\mu$, i.e. on the chemical driving force for
precipitation. If the difference of the chemical potentials exceeds
the stability range of the $\alpha$-phase only slightly, residence
times are rather long (in terms of MC steps) and meaningful averages
(with regard to SRO) can be obtained (see \fig{fig:sro_evo} below for
an example).

As described above, in the early stages of a simulation, some
$\alpha'$-nuclei may exist in the $\alpha$-phase, but they do not yet
exceed the critical size. Fluctuations in the composition eventually
lead to the formation of stable (supercritical)
$\alpha'$-precipitates, which subsequently coarsen. While this
evolution cannot be related to a physical time scale. because atoms do
not move via diffusion but via swaps, it nevertheless represents the
sequence of steps followed during the real evolution of a saturated
$\alpha$-phase. Similarly, SRO develops in the $\alpha$-phase in the
course of the simulation, and reaches equilibrium before the
$\alpha/\alpha'$-microstructure emerges, reflecting the fact that
precipitation is a first order phase transition with a nucleation
barrier, while order-disorder is not.

\subsection{Computational details}
\label{sect:compdetails}

We used samples with $40\times 40\times 40$ conventional bcc unit
cells equivalent to 128,000 atoms. Simulations were carried out for
temperatures between 100 and 900\,K and chemical potential differences
between $-0.45\,\eV/\atom$ and $0.12\,\eV/\atom$. To obtain the data
presented below, the samples were run for a total of $3\times 10^3$ MC
steps/atom. We have found this number to be sufficient to obtain well
converged results for the SRO parameter.
For chemical potential differences which are equivalent to Cr
concentrations below the solubility limit the system equilibrates
quickly. Averages are obtained over all configurations in the
Markov chain after the initial equilibration period.
For values of the chemical potential difference which correspond to
concentrations slightly {\em above} the solubility limit the system
does not immediately separate into two phases due to the nucleation
barrier associated with this process (compare previous
paragraph). Instead one observes a transient state during which the
system equilibrates in the solid solution. Once a supercritical
nucleus of the secondary phase has formed, it grows very quickly and
the system decomposes into two phases. (An example of this behavior is
shown in \fig{fig:sro_evo} below). The existence of this transient
regime allows us to obtain meaningful averages for concentrations
slightly inside the miscibility gap corresponding to oversaturated
solid solutions. (In \fig{fig:sro_evo} this is the case for the first
1000 MC steps). All the data presented in the following sections
(specifically the data shown in \fig{fig:sro_ss} and
\fig{fig:sro_shells}) was obtained in this manner, i.e. as averages
over configurations which correspond to the $\alpha$-phase only.
A detailed description of a simulation which develops from a single to
a two-phase system and which illustrates the transient behavior
referred to above is given in \sect{sect:sro_example}.

The interatomic potential used in this work is taken from
Ref.~\onlinecite{CarCarLop06b}. It is an embedded atom method type
potential based on the Fe potential by Mendelev \etal\
\cite{MenHanSro03} and the Cr potential  by Wallenius
\etal. \cite{WalOlsLag04} To model the Fe--Cr alloy, the potential for
the mixed interaction has been constructed to exactly reproduce the
mixing enthalpy curve obtained from first-principles
calculations.\cite{OlsAbrWal06b}

\section{Results}
\label{sect:results}

\subsection{Chemical potential difference vs. concentration}

\begin{figure}[b]
  \centering
\includegraphics[width=0.92\columnwidth]{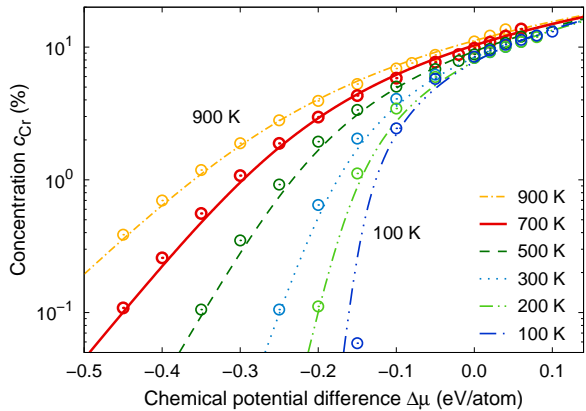}
  \caption{
    (Color online)
    Relation between chemical potential difference and Cr
    concentration in $\alpha$-Fe. The data points were obtained by
    simulation. The lines are fits to \eq{eq:regsol}.
  }
  \label{fig:chempot}
\end{figure}

Figure~\ref{fig:chempot} shows the relation between the equilibrium Cr
concentration $c_{\Cr}$ and the difference of the chemical
potentials $\Delta\mu=\mu_{\Cr}-\mu_{\Fe}$ for different
temperatures. It is found that the dependence cannot be reproduced
using the expression for an ideal solution, $\Delta\mu=\Delta\mu^0+k_B
T\ln c_{\Cr}/(1-c_{\Cr})$. The curves can, however, be reasonably well
fit assuming a regular solution \cite{PorEas92}
\begin{align}
  \Delta\mu
  &= \Delta\mu^0 + \Omega (1-2c_{\Cr})
  + k_B T \ln \frac{c_{\Cr}}{1-c_{\Cr}}
  \label{eq:regsol}
\end{align}
as shown by the lines in \fig{fig:chempot} where $\Delta\mu^0$ and
$\Omega$ were treated as fitting parameters.


\subsection{Short-range ordering in the $\alpha$-phase}
\label{sect:sro_ss}

\begin{figure}
  \centering
\includegraphics[width=0.92\columnwidth]{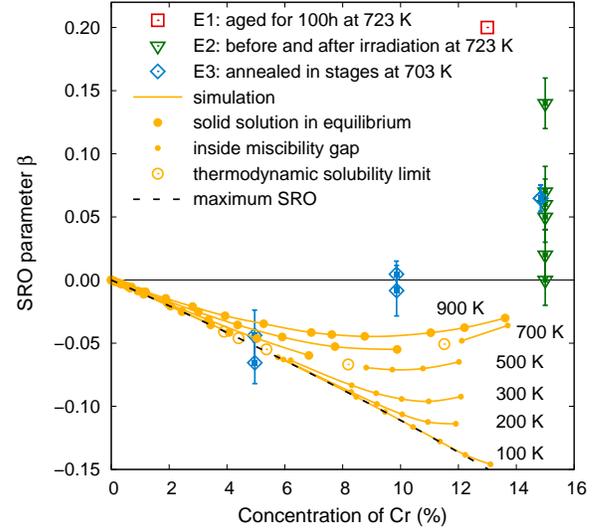}
  \caption{
    (Color online)
    Dependence of the short-range order parameter, $\beta$, defined by
    \eq{eq:beta} on the total Cr concentration. Circles show
    simulation data. Solid lines serve as guide to the eyes. The
    dashed line represents the maximum possible order as given by
    \eq{eq:sro_limit}.
    The solubility limits for each temperature, which is indicated by
    the open circles, were taken from
    Ref.~\onlinecite{CarCarLop06b}. The points which belong to
    concentrations beyond this limit, i.e. inside the miscibility gap,
    are indicated by small filled circles. They have been obtained by
    averaging over configurations of the transient single phase system
    (oversaturated solid solution) which is observed prior to
    nucleation of the $\alpha'$-phase (compare
    \sect{sect:compdetails}).
    E1: experimental data from Ref.~\onlinecite{OvcZviLit76};
    E2: experimental data from Ref.~\onlinecite{OvcGolGus06};
    E3: experimental data from Ref.~\onlinecite{MirHenPar84}.
  }
  \label{fig:sro_ss}
\end{figure}

The results for the bcc SRO parameter $\beta$ defined in
\eq{eq:beta} are shown in \fig{fig:sro_ss}. Several observations can
be made: At small concentrations $\beta$ becomes more negative with
increasing Cr concentration. At low temperature the curves approach
the theoretical lower limit for the SRO parameter (dashed line in
\fig{fig:sro_ss}) given by \eq{eq:sro_limit}. For larger
concentrations the SRO parameter increases, showing a minimum between
8 and 12\%\ Cr the location of which is only weakly dependent on
temperature. We conclude that in the entire compositional range where
only the $\alpha$-phase exists, its SRO is strong. Furthermore, we
conclude that the absolute value of the SRO versus composition has a
maximum at around 10\%\ Cr, quite independent of composition.

\begin{figure}
  \centering
  \includegraphics[width=0.92\columnwidth]{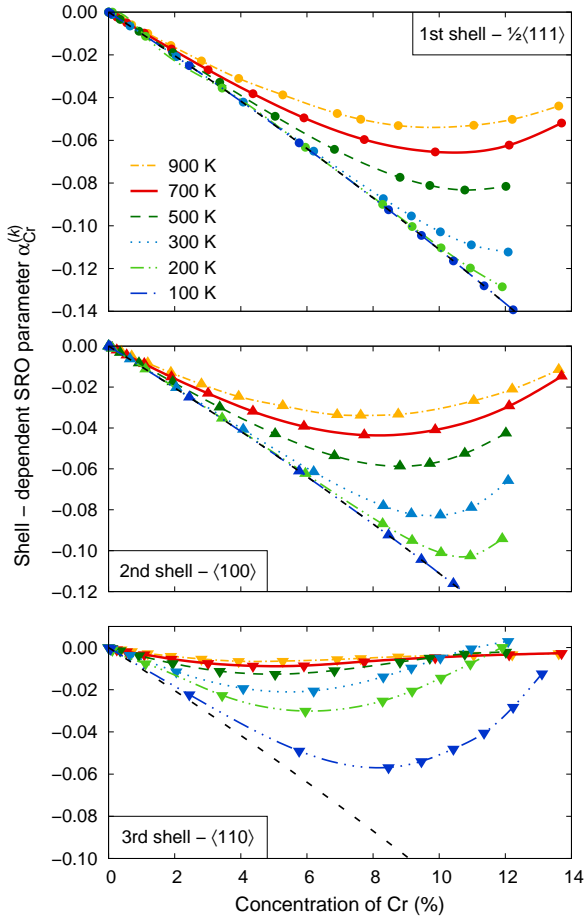}
  \caption{
    (Color online)
    Short-range order parameters for the first three neighbor shells as
    defined in \eq{eq:sro}. The SRO parameter displays a minimum as a function
    of Cr content and becomes less negative with increasing temperature. The
    latter effect is the more pronounced the larger the shell radius.
    The dashed line shows the lower bound for the SRO order given by
    \eq{eq:sro_limit}
    \label{fig:sro_shells}
  }
\end{figure}

The minimum in the curves occurs due to the onset of a saturation
of the Cr distribution: Since the interaction energy between two Cr
atoms in Fe is large and positive,\cite{CarCarKla07} the lowest energy
configurations maximize the separation between the Cr atoms. This is
the source of the ordering tendency in this system which attempts
to maximize the number of non-alike neighbors,
i.e. $Z_{\Fe}^{(k)}/Z_{\tot}^{(k)}$ in \eq{eq:sro} approaches one.

As the Cr concentration increases, the fraction of configurations
which maximize the Cr--Cr separation decreases. At the same time the
entropy gained by randomizing the system increases significantly since
the total number of possible configurations increases rapidly.
The SRO is therefore both strongly temperature and concentration
dependent. In the limit of zero temperature, i.e. in the absence of
entropic effects, the SRO parameter $\beta$ approaches its lower bound
given by \eq{eq:sro_limit}. At high temperatures it becomes
less negative since the entropy favors randomization of the system.
As the Cr concentration increases the entropy gained by
randomization eventually exceeds the energy gained by ordering causing
the minimum in the SRO parameter curves.

This effect is further illustrated in \fig{fig:sro_shells} which shows
the SRO parameters $\alpha_{\Cr}^{(k)}$ for the first three neighbor shells.
For the first shell the calculated SRO parameter at low temperatures
follows closely the theoretical limit [\eq{eq:sro_limit}]
corresponding to a situation in which the first shell around any Cr
atom is exclusively filled with Fe atoms. With increasing temperature
the point at which the SRO parameter, $\alpha_{\Cr}^{(1)}$, deviates
from the theoretical limit is shifted towards smaller
concentrations. For the second and third shells this deviation occurs
at yet lower temperatures and concentrations.
This behavior is correlated with the interaction energy {\em vs.}
separation curves given in Ref.~\onlinecite{CarCarKla07} which show a
significant drop of the Cr repulsion energy with separation.

As mentioned above the SRO parameter $\beta$ for larger Cr concentrations
becomes less negative with increasing temperature. However, for the
interaction model used in the present work it does not become zero
before melting, which would indicate a completely random solution. It
should be pointed out that this observation is a property of this
model and is not necessarily related to the behavior of the real
alloy, which undergoes both magnetic and structural transitions
(ferromagnetic--paramagnetic, $\alpha-\gamma-\delta$) before melting,
transitions that are not described by the present empirical
potential.

Figure~\ref{fig:sro_ss} includes several experimental data
points.\footnote{
  Some data points were directly extracted from figures in the
  original references using the software \textsc{g3data}
  (Ref.~\onlinecite{g3data}).}
They indicate a transition of the SRO parameter through a minimum and
subsequently an increase up to positive values. The occurrence of a
minimum is seemingly similar to the simulation results but the two
minima have different physical origins:
The experimentally determined SRO represents an average over the
entire sample, eventually including both the $\alpha$-phase and
$\alpha'$-precipitates. In contrast, the simulation data shown
in \fig{fig:sro_ss} have been explicitly obtained for the
$\alpha$-phase only (compare \sect{sect:compdetails}) and the
occurrence of a minimum is an intrinsic property of this solid
solution (as discussed above). Thus, in order to explain the relation
between the experimentally measured SRO which involves the entire
sample and its values in the $\alpha$ and $\alpha'$-phases, we must
first understand the contribution of the $\alpha'$-precipitates to the
measured SRO parameter.

\subsection{Bimodal distribution of
  the SRO parameters in the two-phase region}
\label{sect:sro_example}

\begin{figure}
  \centering
\includegraphics[width=0.92\columnwidth]{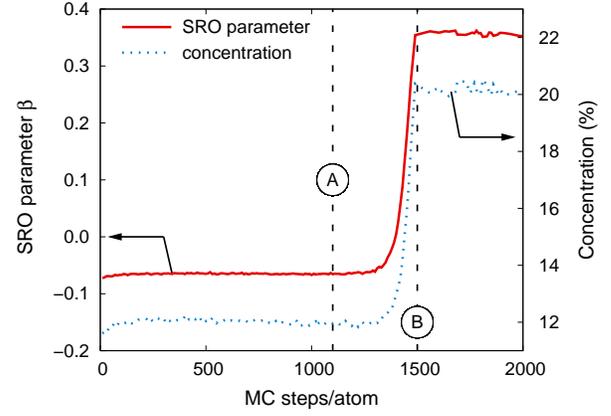}
  \caption{
    (Color online)
    Evolution of the average concentration and SRO parameter during
    the course of a simulation ($T=500\,\K$,
    $\Delta\mu=0.06\,\eV/\atom$). The points A and B refer to the
    configurations shown in Figures~\ref{fig:sro_dist} and
    \ref{fig:confs}.
  }
  \label{fig:sro_evo}
\end{figure}

\begin{figure}[b]
  \centering
\includegraphics[width=0.92\columnwidth]{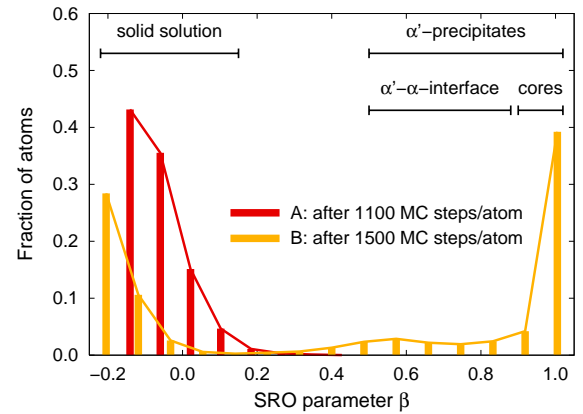}
  \caption{
    (Color online)
    Number distribution of SRO parameter before (A) and after (B)
    the formation of supercritically sized $\alpha'$-precipitates. The
    corresponding configurations are shown in \fig{fig:confs}. The simulation
    was carried out at 500\,K using a chemical potential difference of
    $\Delta\mu=0.06\,\eV/\atom$.
  }
  \label{fig:sro_dist}
\end{figure}

\begin{figure}
  \centering
\includegraphics[width=0.72\columnwidth]{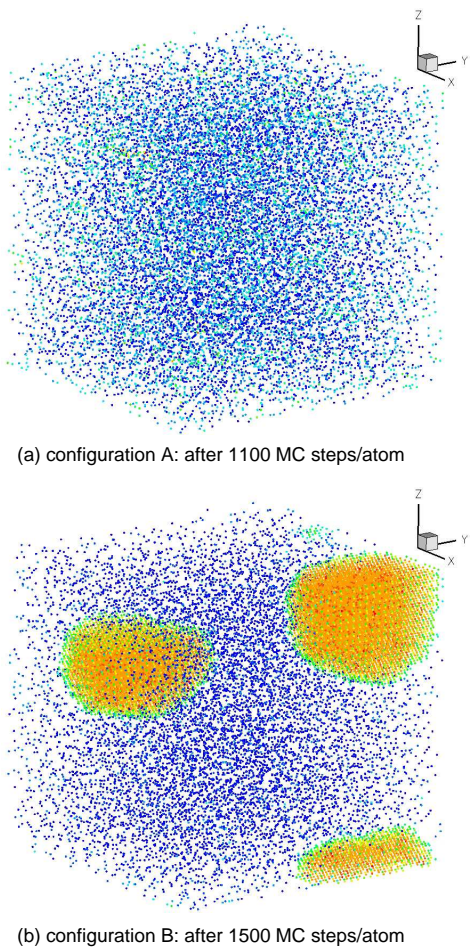}
  \caption{
    (Color online)
    Atomic configurations before (a) and after (b) the formation of
    supercritical $\alpha'$-precipitates. Only Cr atoms are
    shown with a color coding based on their SRO parameter:
    blue/green (dark gray) -- negative or small positive SRO parameter
    (Fe-rich solid solution, $\alpha$-phase),
    yellow/red (light gray) -- positive SRO parameter
    ($\alpha'$-precipitates, $\alpha/\alpha'$-interface).
    The simulation was carried out at 500\,K using a chemical
    potential difference of $\Delta\mu=0.06\,\eV/\atom$.
  }
  \label{fig:confs}
\end{figure}

In order to illustrate the contributions of the $\alpha$-phase and
$\alpha'$-precipitates to the SRO parameter, we consider one
particular simulation ($T=500\,\K$, $\Delta\mu=0.06\,\eV/\atom$). At
this temperature, the chemical potential difference chosen implies
that without constraints on the total concentration, the system in
equilibrium would be in the $\alpha'$-phase, whereas the constraint
keeps the system inside the two-phase region. This particular
simulation is suitable for illustrating the transition from a single
to a two-phase system during the course of the MC simulation:
The initial sample contains Fe atoms only and the SRO parameter
$\beta$ measured over the entire sample is zero. During the very first
steps of the simulation the number of Cr atoms increases
sharply. Already after about 200 MC steps/atom the concentration of Cr
settles to a value of about $12\%$ and the SRO parameter $\beta$
adopts a value of about $-0.07$ as shown in \fig{fig:sro_evo}. In this
stage the entire sample is a Fe-rich saturated homogeneous solid
solution.

Figure~\ref{fig:sro_dist} shows histograms of the number distribution
of SRO parameters obtained by averaging over every Cr atom in the
sample after 1100 MC steps per atom (point A in \fig{fig:sro_evo})
which is a representative configuration for the Fe-rich homogeneous
solid solution (pure $\alpha$-phase). There is one pronounced peak on
the negative side equivalent to the SRO of the $\alpha$-phase. Note
that the range of SRO parameter also includes some positive
values. This is related to the saturation of the $\alpha$-phase, which
in equilibrium at this temperature has a solubility of about 8\%\
Cr\cite{CarCarLop06b}, and to the formation of unstable (subcritically
sized) $\alpha'$-nuclei. For further illustration the configuration is
shown in \fig{fig:confs}(a) clearly indicating the absence of
precipitation.

The system remains in this metastable state up to about 1300 MC
steps per atom. Then it undergoes a rapid transition as indicated
both by the SRO parameter (average over the entire sample) and the
concentration. At this point supercritically sized $\alpha'$-precipitates have
formed which subsequently grow. Without constraints their size would
increase until the {\em total} concentration of Fe reaches the
the value corresponding to the given chemical potential difference on
the Cr-rich side of the phase diagram. The final Fe concentration will
therefore assume some value lower than the solubility of Fe in Cr. In
the present simulations, the Cr concentration is however enforced to
be about $20\%$ or lower. The distribution of SRO parameters shown after
1500 MC steps/atom (point B in \fig{fig:sro_evo}) in
\fig{fig:sro_dist} now clearly displays two distinct peaks which
result from the $\alpha$-phase ($\beta\lesssim 0.2$) and the
$\alpha'$-precipitates ($\beta\approx 1$), respectively (compare
\sect{sect:discussion}).

The resulting SRO parameter averaged over the entire sample is about
$0.35$. Note that the SRO parameter of the Cr atoms in the
$\alpha$-phase drops after the occurrence of
$\alpha'$-precipitates. This is mostly due to the higher global Cr
concentration which enters in \eq{eq:sro}. It is therefore an effect
of the concentration dependence entering the definition of the SRO
parameter, not an indication for an actual higher degree of order in
the $\alpha$-phase. At the same time the
concentration of Cr in the $\alpha$-phase returns to the solubility
limit. This behavior corresponds to the curves in
\fig{fig:sro_ss}. The atomic configuration after 1500 MC steps/atom is
visualized in \fig{fig:confs}(b) which clearly shows the formation of
precipitates and the emergence of a two-phase mixture. The color scale
furthermore illustrates the intermediate SRO parameter obtained for
atoms at the $\alpha/\alpha'$-interface.

This example clearly illustrates the transition from a Fe-rich
saturated solid solution to a two-phase system. It demonstrates how in
this range of Cr concentrations the local SRO parameter can be used to
identify atoms as being either part of the $\alpha$ or $\alpha'$-phase
(see in particular \fig{fig:confs}(b)). Furthermore, it elucidates the 
different contributions which lead to a given measurable SRO parameter
and how the latter can be utilized to detect the formation of 
$\alpha'$-precipitates. In the following section, these observations
will be generalized and combined to obtain a general expression for
the average SRO parameter as a function of Cr concentration.

\begin{figure}
  \centering
  \includegraphics[width=0.92\columnwidth]{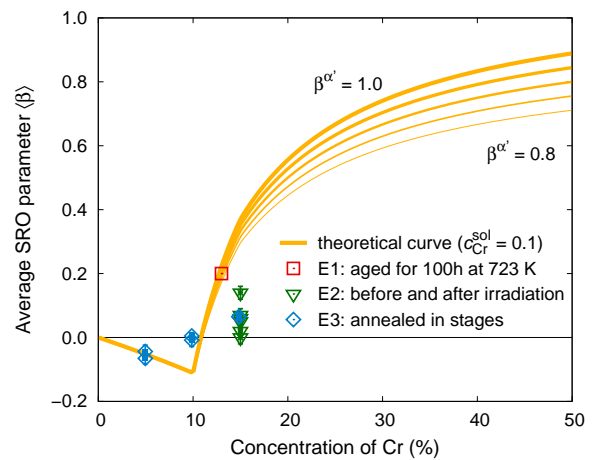}
  \caption{
    (Color online)
    Theoretical prediction for the average SRO parameter $\left<\beta\right>$
    as given by \eq{eq:sro_avg} using \eq{eq:sro_limit} to express
    the concentration dependence of the SRO parameter in the
    $\alpha$-phase. Experimental references as given in the caption of
    \fig{fig:sro_ss}.
    The set of lines between $\beta^{\alpha'}=0.8$ and $\beta^{\alpha'}=1.0$
    correspond to different values of $\beta^{\alpha'}$ schematically
    indicating the effect of coarsening of the $\alpha'$-precipitates.
    \label{fig:sro_avg}
  }
\end{figure}

\section{Discussion}
\label{sect:discussion}

\subsection{Interpretation of the SRO inversion}
\label{sect:discussion_limit}

The results presented above demonstrate that particular care must be
exercised when interpreting the SRO parameter in the case of a
two-phase mixture. In the following, we focus on the case of an
Fe-rich Fe--Cr alloy at low temperature.

While the SRO parameter for Cr atoms in the $\alpha$-phase is
typically small and negative, the SRO parameter for Cr atoms in the
$\alpha'$-phase is close to one. Since experimentally the SRO is
obtained over the entire sample, the measured value is an average over
all types of local short-range order. This average is the weighted sum
of the contributions from the $\alpha$ and $\alpha'$-phases as well as
the $\alpha/\alpha'$-interface. However, for simplicity in what
follows we subsume the $\alpha/\alpha'$-interface contribution into
the contribution from $\alpha'$, i.e. we treat the SRO parameter
of the $\alpha'$-phase as if it were dependent on the size of the
precipitates. This assumption greatly simplified the following
treatment and the SRO parameter averaged over the entire sample can be
expressed as follows
\begin{align}
  \left<\beta\right>
  =    \phi^{\alpha}     \beta^{\alpha}[c_{\Cr}^{\tot}]
  +    (1-\phi^{\alpha}) \beta^{\alpha'}[c_{\Cr}^{\tot}]
  \label{eq:sro_avg}
\end{align}
where $c_{\Cr}^{\tot}$ denotes the total concentration of Cr in the
system and $\beta^j$ is the SRO parameter for the Cr atoms belonging
to the $\alpha$ or the $\alpha'$-phase, respectively. The fraction of
Cr atoms in the $\alpha$-phase, $\phi^{\alpha}$, can be derived as
shown in the appendix.

According to \eq{eq:sro_avg} below the solubility limit
$\left<\beta\right>$ equals the SRO parameter of the $\alpha$-phase,
the concentration and temperature dependence of which have been
discussed in \sect{sect:sro_ss}.

We now need to discuss the terms in the equation for
$\left<\beta\right>$ which determine the average SRO parameter inside
the miscibility gap:
The thermodynamic boundary conditions require the Cr concentration in
the $\alpha$-phase to be equal to the solubility of Cr in Fe and
reciprocally for the $\alpha'$-phase. Due to these constraints the
proportions of $\alpha$ and $\alpha'$ depend only on the total Cr
concentration (compare appendix). Furthermore, since the empirical
potential \footnote{
  The potential used in the present study has been parameterized to
  exactly reproduce the mixing enthalpy obtained from 0\,K DFT
  calculations (Ref.~\onlinecite{OlsAbrWal06b}). The mixing curve was
  calculated with respect to the 0\,K reference states (ferromagnetic
  Fe, anti-ferromagnetic Cr). Since the real system undergoes several
  temperature driven transitions the reference states change and the
  accordingly the mixing enthalpy. At present this change is, however,
  not captured by the potential, which causes an underestimation of
  the solubility of Fe in the $\alpha'$-phase at elevated
  temperatures.  This shortcoming does, however, not affect the main
  conclusions of the present work.
} used in this study underestimates the solubility of Fe in
Cr, the $\alpha'$-precipitates contain nearly 100\%\ Cr. The
solubility also affects the SRO parameter for the $\alpha'$-phase
since it correlates with the ratio
$Z_{\Fe}^{(k)}/Z{\tot}^{(k)}$. However, for Fe-rich conditions this
can be safely neglected -- again -- because of the small
solubility. This conclusion is in line with the observations described
in \sect{sect:sro_example}.

In \eq{eq:sro_avg} we have subsumed the contribution of the
$\alpha/\alpha'$-interface into $\beta^{\alpha'}$. As shown for
example in Figs.~\ref{fig:sro_dist} and \ref{fig:confs}(b) the SRO
parameter for Cr atoms at the interface is significantly smaller than
for atoms in the interior. Therefore, both $\beta^{\alpha'}$ and
$\left<\beta\right>$ depend on the surface to volume ratio of the
precipitates. The maximum value for $\beta^{\alpha'}$ is obtained for
the limit of a completely coarsened system for which the contribution
of the $\alpha/\alpha'$-interface is practically zero
($\beta^{\alpha'}\approx 1$). At the onset of precipitation the size of
the $\alpha'$-precipitates is, however, small and the surface to volume ratio
is large. The effective $\beta^{\alpha'}$ is therefore reduced with respect to
the fully decomposed system ($\beta^{\alpha'}\approx 0.8$). Hence,
depending on the kinetics of the system $\beta^{\alpha'}$ and
$\left<\beta\right>$ can assume a range of values. Both are expected
to grow gradually as the system coarsens and gets closer and closer to
equilibrium. This is illustrated in \fig{fig:sro_avg} as described in
the following.

If one uses the theoretical lower limit for $\beta^{\alpha}$, which as
shown in \fig{fig:sro_ss} represents the low temperature limit of the
simulation data for the $\alpha$-phase, one obtains an entirely
analytic expression for $\left<\beta\right>$.
The resulting curves for different values of $\beta^{\alpha'}$ are
shown in \fig{fig:sro_avg} where the solubility of Cr has been set to
$c_{\Cr}^{\sol}=10\%$ roughly corresponding to the experimental
solubility at this temperature. All curves are negative for small Cr
concentrations, then pass through zero and eventually approach
saturation. The transition from negative to positive values is the
hallmark of Cr precipitation and rather sharp.
As discussed above at a given temperature the only variable is the
SRO parameter for the $\alpha'$-precipitates which depends on their
size distribution. The most extremal value is $\beta^{\alpha'}=1$
which is achieved if the number of $\alpha/\alpha'$-interface
atoms is negligible with respect to the number of atoms in
$\alpha'$-precipitates. The resulting curve is shown by the most upper
line in \fig{fig:sro_avg}. For smaller $\alpha'$-precipitates the
SRO parameter $\beta^{\alpha'}$ is smaller than one. The resulting
SRO parameter, $\left<\beta\right>$, is shown in \fig{fig:sro_avg} for
several values between $\beta^{\alpha'}=0.8$ and $0.95$.
Upon aging the average SRO parameter at a given concentration and
temperature thus will gradually grow from the lower to the upper
curves in \fig{fig:sro_avg}.
It is apparent from the figure that the effect of a reduced
$\beta^{\alpha'}$ becomes notable for $c_{\Cr}^{\tot}\gtrsim 0.2$.

It should be stressed that the interpretation of the evolution of the average
SRO parameter $\left<\beta\right>$ upon aging is solely based on a gradual
approach to equilibrium, i.e. the coarsening of the precipitates. This
conclusion is independent of our MC simulations.

\subsection{Comparison with experiments}

On the basis of the foregoing analysis, it is now possible to carry
out a quantitative comparison with the experimental data points
included in Figures \ref{fig:sro_ss}, \ref{fig:sro_avg} and
\ref{fig:sro_temp}. Since all experiments considered here correspond
to temperatures close to 700\,K, we focus on the SRO curve calculated
at this temperature.

The experimental data point obtained at $5\%$ Cr corresponds to a
Fe-rich homogeneous solid solution and the simulation data for the
$\alpha$-phase (\fig{fig:sro_ss}) as well as the predicted average SRO
parameter $\left<\beta\right>$ (\fig{fig:sro_avg}) are in very good
agreement.

We suggest that at $10\%$ Cr $\alpha'$-precipitates are formed in the
experimental samples which drive the SRO parameter upwards leading to
an average SRO parameter value close to zero. This trend is reproduced
by the theoretical curve which yields a transition through zero close
to the experimental data point.
The comparison with the calculated curves shows that the inversion of
the SRO is very steep and is shifted to slightly larger concentration
with respect to the solubility limit.

There are several data points at concentrations close to 15\%\ which
are scattered between $\left<\beta\right>=0.0$ and 0.2:
The measurements in Ref.~\onlinecite{OvcZviLit76} (E1 in
\fig{fig:sro_temp}) were carried out on a sample aged for 100 hours at
723\,K yielding the largest value for the SRO parameter among the
experimental data points. The sample measured in
Ref.~\onlinecite{MirHenPar84} (E3 in \fig{fig:sro_temp}) was subjected
to isochronal annealing steps at 703\,K. In
Ref.~\onlinecite{OvcGolGus06} (E2 in \fig{fig:sro_temp}) the authors
carried out measurements on different samples before and after
irradiation. All samples were initially quenched from 1473\,K and
subsequently annealed/irradiated at 723\,K. Irradiation experiments
were carried out using both Ar$^+$ and Fe$^+$ ions and exposure times
ranging from 3 to 30\,min. Within the experimental error bars the SRO
parameter was zero for the initial sample. It was observed to increase
with exposure time in a series of bombardments with Fe$^+$. The
largest SRO was obtained after extended irradiation with Ar$^+$ ions.

Collecting the experimental data and using the above analysis, the
following interpretation for the observations at higher Cr
concentrations emerges:
(1)
The samples initially obtained by quenching from high temperature show
a very small SRO parameter close to zero. This suggests that during
rapid quenching from elevated temperatures both the ordering tendency
and more importantly the precipitation and growth of
$\alpha'$-precipitates are suppressed (lowest data point from E2).
(2) On the other hand, the most extensively aged samples show a rather
large SRO parameter which is close to the values predicted by the
theoretical analysis presented above (E1). This suggests that the
$\alpha'$-precipitates in these samples have already reached rather
large dimensions (diminishing the contribution of the
$\alpha/\alpha'$-interface).
(3)
Samples subjected to energetic ion beams showed an increase of the
average SRO parameter with the intensity of the
irradiation, reflecting radiation induced precipitation.

\subsection{Temperature dependence}

\begin{figure}[b]
  \centering
  \includegraphics[width=0.92\columnwidth]{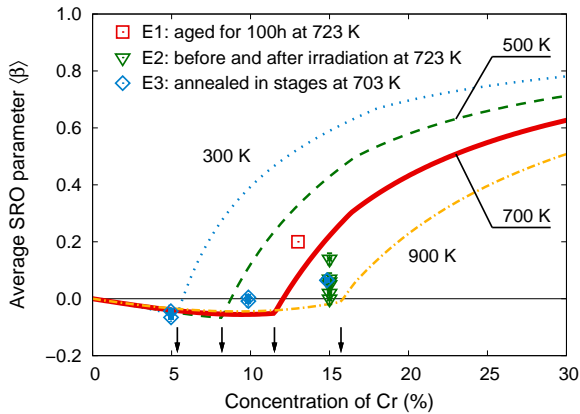}
  \caption{
    (Color online)
    Theoretical prediction for the average SRO parameter based on
    \eq{eq:sro_avg} using $\beta^{\alpha'}=0.9$ and the simulation data from
    \fig{fig:sro_ss} for the SRO parameter in the $\alpha$-phase. The
    temperature dependence of the solubility was taken from
    Ref.~\onlinecite{CarCarLop06b} and is indicated by the small
    arrows at the bottom of the figure.
    \label{fig:sro_temp}
  }
\end{figure}

The shape of the SRO curves and the transition through zero are
affected by temperature because of the $T$-dependence of both the SRO
in the $\alpha$-phase (\fig{fig:sro_ss}) and the solubility of Cr in
$\alpha$-Fe.\cite{CarCarLop06b} If these effects are taken into
account one obtains the curves shown in \fig{fig:sro_temp}. The
increase of the solubility with temperature \cite{CarCarLop06b} shifts the
transition of the SRO parameter through zero to larger concentrations, whereas
the $T$-dependence of the SRO in the $\alpha$-phase (\fig{fig:sro_ss}) causes
a slight upward shift of the average SRO. Note that the transition through
zero is correlated with but not identical to the solubility of Cr.

\section{Conclusions}
\label{sect:conclusions}

In this paper we have investigated the short-range order (SRO) in
Fe-rich Fe--Cr alloys using atomistic simulations based on an
empirical potential description of the alloy. For low temperatures
and small concentrations the SRO in the $\alpha$-phase is found to be
close to the theoretical maximum possible SRO. As temperature
increases SRO is reduced due to entropic effects. For somewhat larger
Cr concentrations the SRO curves go through a minimum, which occurs close to
the solubility limit. The occurrence of a minimum reflects the competition
between energy induced ordering and entropy-driven randomization.  It is,
however, important to stress that directly at the solubility the SRO is still
rather strong.

In the two-phase region the experimentally measured SRO parameter is
a mixture of both the SRO in the $\alpha$-phase and the
$\alpha'$-precipitates. The MC simulations show that the two
contributions can be clearly separated. When using a weighted average
of these contributions the average SRO parameter is predicted as a
function of concentration in good agreement with experiment.
The MC simulations in conjunction with this simple model demonstrate that the
inversion of the SRO parameter observed experimentally is a result of the
formation of stable (supercritical) $\alpha'$-precipitates. It is {\em not}
related to the loss of SRO in the $\alpha$-phase, since directly at the
solubility limit there is still an appreciable degree of SRO. The model has
also been used to investigate the effect of temperature on the SRO parameter
which shows the temperature dependence of the solubility of Cr in Fe to be the
dominating factor.

The results reported here are anticipated to support future research
in at least two aspects: (1) As indicated in the introduction the
SRO affects the mobility of dislocations. The present paper
therefore constitutes a basis for a detailed atomistic study of the
mobility of dislocations in Fe--Cr alloys following a strategy
similar to Ref.~\onlinecite{MarCar06}. (2) The present paper
clarifies the contributions to the average SRO parameter in a
two-phase system and introduces a simple predictive model. It
thereby provides a better understanding of previous experimental
results and provides the basis for a more elaborate interpretation
of future experiments.

\begin{acknowledgments}
This work was performed under the auspices of the U.S. Department of
Energy by the University of California, Lawrence Livermore National
Laboratory under Contract No. DE-AC52-07NA27344 with support from the
Laboratory Directed Research and Development Program. Generous grants
of computer time through the National Energy Research Scientific
Computing Center at Lawrence Berkeley National Laboratory are
gratefully acknowledged.
\end{acknowledgments}

\vspace{-14pt}
\appendix*
\section{}

The total number of Fe/Cr atoms is simply the sum of the number of
Fe/Cr atoms in the $\alpha$-phase and the $\alpha'$-precipitates
\begin{align}
  N_{\Fe}^{\tot} = N_{\Fe}^{\alpha} + N_{\Fe}^{\alpha'}
  \quad
  \text{and}
  \quad
  N_{\Cr}^{\tot} = N_{\Cr}^{\alpha} + N_{\Cr}^{\alpha'}. \label{eq:eq1}
\end{align}
In equilibrium the concentration of Cr in the Fe-matrix is given
the maximum solubility, $c_{\Cr}^{\sol}$. It can be expressed by the
number of Fe and Cr atoms in the Fe-rich solid solution
$c_{\Cr}^{\sol} = N_{\Cr}^{\alpha}/(N_{\Fe}^{\alpha}+N_{\Cr}^{\alpha})$
which can be rearranged to give
\begin{align}
  N_{\Cr}^{\alpha} = x_{\Cr}^{\sol} N_{\Fe}^{\alpha}.
  \quad
  \text{with}
  \quad
  x_i^j = c_i^j/(1-c_i^j).
  \label{eq:eq2}
  \\\phantom{x}\nonumber
\end{align}
Similarly, the number of Fe atoms in the $\alpha'$-precipitates is
connected to the equilibrium solubility of Fe in Cr
\begin{align}
  N_{\Fe}^{\alpha'} &= x_{\Fe}^{\sol} N_{\Cr}^{\alpha'}.
  \label{eq:eq3}
\end{align}
Combining equations (\ref{eq:eq1}--\ref{eq:eq3}) one obtains
\begin{align}
  N_{\Cr}^{\alpha}
  &= x_{\Cr}^{\sol} (N_{\Fe}^{\tot} - N_{\Fe}^{\alpha'})
  = x_{\Cr}^{\sol} (N_{\Fe}^{\tot} - x_{\Fe}^{\sol} N_{\Cr}^{\alpha'}).
\end{align}
Using $N_{\Fe}^{\tot}=x_{\Cr}^{\tot} N_{\Cr}^{\tot}$ and rearranging
finally yields
\begin{align}
  \frac{N_{\Cr}^{\alpha}}{N_{\Cr}^{\tot}}
  &=
  \frac{
    1/x_{\Cr}^{\tot} - x_{\Fe}^{\sol}
  }{
    1/x_{\Cr}^{\sol} - x_{\Fe}^{\sol}
  }
  = \phi^{\alpha}.
\end{align}
This expression gives the fraction of Cr atoms in the $\alpha$-phase
as a function of the total Cr concentration within the miscibility
gap. For Cr concentrations below $c_{\Cr}^{\sol}$ (above
$c_{\Fe}^{\sol}$) $\phi^{\alpha}$ is one (zero). In summary,
$\phi_{\alpha}$ can be written as
\begin{align}
  \phi^{\alpha}
  =
\begin{cases}
  1
  & c_{\Cr}^{\tot}<c_{\Cr}^{\sol} \\
  {
    \displaystyle
    \frac{
      1/x_{\Cr}^{\tot} - x_{\Fe}^{\sol}
    }{
      1/x_{\Cr}^{\sol} - x_{\Fe}^{\sol}
    }
  }
  & c_{\Cr}^{\sol} < c_{\Cr}^{\tot} < c_{\Fe}^{\sol} \\
  0
  & c_{\Cr}^{\tot}>c_{\Fe}^{\sol}
  \end{cases}.
\end{align}

\end{document}